\pdfoutput=1

\documentclass[12pt,a4paper]{article}

\usepackage{feynmp-auto}
\usepackage[compat=1.1.0]{tikz-feynman}
\usepackage{subcaption}
\usepackage{ifthen} 
\newboolean{pdflatex}
\setboolean{pdflatex}{true} 

\newboolean{articletitles}
\setboolean{articletitles}{true} 

\newboolean{uprightparticles}
\setboolean{uprightparticles}{false} 

\def\paperauthors{LHCb collaboration} 
\def\paperasciititle{Observation of the Xic0 -> pK- decay and measurement of its decay asymmetry}
\def\papertitle{Observation of the \decay{\Xicz}{\proton\Km} decay and measurement of its decay asymmetry} 
\def\paperkeywords{{High Energy Physics}, {LHCb}} 
\def\papercopyright{\the\year\ CERN for the benefit of the LHCb collaboration}
\def\paperlicence{CC BY 4.0 licence}
\def\paperlicenceurl{https://creativecommons.org/licenses/by/4.0/}

\newif\ifEnableSectionTOCLinks
\EnableSectionTOCLinksfalse 

\usepackage[top=1in, bottom=1.25in, left=1in, right=1in]{geometry}

\usepackage{microtype}
\usepackage{lineno}  
\usepackage{xspace} 
\usepackage{caption} 

\usepackage{graphicx}  
\usepackage{color}
\usepackage{colortbl}
\graphicspath{{./figs/}} 

\usepackage{amsmath} 
\usepackage{amssymb}
\usepackage{amsfonts}
\usepackage{upgreek} 

\usepackage[normalem]{ulem} 

\newcommand*\patchAmsMathEnvironmentForLineno[1]{%
\expandafter\let\csname old#1\expandafter\endcsname\csname #1\endcsname
\expandafter\let\csname oldend#1\expandafter\endcsname\csname
end#1\endcsname
 \renewenvironment{#1}%
   {\linenomath\csname old#1\endcsname}%
   {\csname oldend#1\endcsname\endlinenomath}%
}
\newcommand*\patchBothAmsMathEnvironmentsForLineno[1]{%
  \patchAmsMathEnvironmentForLineno{#1}%
  \patchAmsMathEnvironmentForLineno{#1*}%
}
\AtBeginDocument{%
\patchBothAmsMathEnvironmentsForLineno{equation}%
\patchBothAmsMathEnvironmentsForLineno{align}%
\patchBothAmsMathEnvironmentsForLineno{flalign}%
\patchBothAmsMathEnvironmentsForLineno{alignat}%
\patchBothAmsMathEnvironmentsForLineno{gather}%
\patchBothAmsMathEnvironmentsForLineno{multline}%
\patchBothAmsMathEnvironmentsForLineno{eqnarray}%
}

\usepackage[pdftex,
            pdfauthor={\paperauthors},
            pdftitle={\paperasciititle},
            pdfkeywords={\paperkeywords}]{hyperref}
\usepackage{hyperxmp}
\hypersetup{
    pdfcopyright={Copyright (C) \papercopyright},
    pdflicenseurl={\paperlicenceurl}
}

\usepackage[colorinlistoftodos,textsize=scriptsize]{todonotes}

\usepackage[bottom,flushmargin,hang,multiple]{footmisc}

\usepackage[all]{hypcap} 

\usepackage{xspace} 
\usepackage{upgreek}

\def\lhcb   {\mbox{LHCb}\xspace}

\def\belle  {\mbox{Belle}\xspace}
\def\belletwo {\mbox{Belle~II}\xspace}
\def\besiii {\mbox{BESIII}\xspace}

\def\MagUp {\mbox{\em Mag\kern -0.05em Up}\xspace}

\ifdefined\ifuprightparticles
\else
\newboolean{uprightparticles}
\setboolean{uprightparticles}{false} 
\fi

\ifthenelse{\boolean{uprightparticles}}%
{

 \def\Ppi         {\ensuremath{\uppi}\xspace}

 \def\PDelta      {\ensuremath{\Delta}\xspace}                 
 \def\PXi         {\ensuremath{\Xi}\xspace}                 
 \def\PLambda     {\ensuremath{\Lambda}\xspace}                 
 \def\PSigma      {\ensuremath{\Sigma}\xspace}                 
 \def\POmega      {\ensuremath{\Omega}\xspace}                 
 \def\PUpsilon    {\ensuremath{\Upsilon}\xspace}
 \let\oldPi\Pi
 \def\PPi         {\ensuremath{\oldPi}\xspace}

 \def\PB      {\ensuremath{\mathrm{B}}\xspace}                 
 \def\PD      {\ensuremath{\mathrm{D}}\xspace}                 

 \def\PK      {\ensuremath{\mathrm{K}}\xspace}                 
 \def\Pb      {\ensuremath{\mathrm{b}}\xspace}                 
 \def\Pc      {\ensuremath{\mathrm{c}}\xspace}

 \def\Pp      {\ensuremath{\mathrm{p}}\xspace}                 

 \def\Ps      {\ensuremath{\mathrm{s}}\xspace}

 \def\thebaroffset{0.0em}
}
{

 \def\Ppi         {\ensuremath{\pi}\xspace}

 \mathchardef\PDelta="7101
 \mathchardef\PXi="7104
 \mathchardef\PLambda="7103
 \mathchardef\PSigma="7106
 \mathchardef\POmega="710A
 \mathchardef\PUpsilon="7107
 \mathchardef\PPi="7105
 \def\PB      {\ensuremath{B}\xspace}                 
 \def\PD      {\ensuremath{D}\xspace}                 

 \def\PK      {\ensuremath{K}\xspace}                 
 \def\Pb      {\ensuremath{b}\xspace}                 
 \def\Pc      {\ensuremath{c}\xspace}

 \def\Pp      {\ensuremath{p}\xspace}                 

 \def\Ps      {\ensuremath{s}\xspace}

 \def\thebaroffset{0.18em}
}
\newcommand{\offsetoverline}[2][\thebaroffset]{\kern #1\overline{\kern -#1 #2}}%

\makeatletter
\ifcase \@ptsize \relax
  \newcommand{\miniscule}{\@setfontsize\miniscule{4}{5}}
\or
  \newcommand{\miniscule}{\@setfontsize\miniscule{5}{6}}
\or
  \newcommand{\miniscule}{\@setfontsize\miniscule{5}{6}}
\fi
\makeatother

\DeclareRobustCommand{\optbar}[1]{\shortstack{{\miniscule (\rule[.5ex]{1.25em}{.18mm})}
  \\ [-.7ex] $#1$}}

\def\squark    {{\ensuremath{\Ps}}\xspace}

\def\cquark    {{\ensuremath{\Pc}}\xspace}

\def\bquark    {{\ensuremath{\Pb}}\xspace}

\def\pion   {{\ensuremath{\Ppi}}\xspace}
\def\piz    {{\ensuremath{\pion^0}}\xspace}
\def\pip    {{\ensuremath{\pion^+}}\xspace}
\def\pim    {{\ensuremath{\pion^-}}\xspace}

\def\kaon    {{\ensuremath{\PK}}\xspace}

\def\KorKbar {\kern \thebaroffset\optbar{\kern -\thebaroffset \PK}{}\xspace}

\def\Kp      {{\ensuremath{\kaon^+}}\xspace}
\def\Km      {{\ensuremath{\kaon^-}}\xspace}

\def\D       {{\ensuremath{\PD}}\xspace}

\def\DorDbar {\kern \thebaroffset\optbar{\kern -\thebaroffset \PD}\xspace}
\def\Dz      {{\ensuremath{\D^0}}\xspace}

\def\Dp      {{\ensuremath{\D^+}}\xspace}
\def\Dm      {{\ensuremath{\D^-}}\xspace}

\def\DpDm    {\ensuremath{\Dp {\kern -0.16em \Dm}}\xspace}

\def\B       {{\ensuremath{\PB}}\xspace}

\def\BorBbar {\kern \thebaroffset\optbar{\kern -\thebaroffset \PB}\xspace}

\def\Bd      {{\ensuremath{\B^0}}\xspace}

\def\BdorBdbar {\kern \thebaroffset\optbar{\kern -\thebaroffset \Bd}\xspace}

\def\Bub     {{\ensuremath{\B^-}}\xspace}

\def\Bm      {{\ensuremath{\Bub}}\xspace}

\def\Bs      {{\ensuremath{\B^0_\squark}}\xspace}

\def\BsorBsbar {\kern \thebaroffset\optbar{\kern -\thebaroffset \Bs}\xspace}

\def\Y#1S{\ensuremath{\PUpsilon{(#1S)}}\xspace}

\def\proton      {{\ensuremath{\Pp}}\xspace}

\def\Lz          {{\ensuremath{\PLambda}}\xspace}

\def\LorLbar     {\kern \thebaroffset\optbar{\kern -\thebaroffset \PLambda}\xspace}

\def\Xires       {{\ensuremath{\PXi}}\xspace}
\def\Xiz         {{\ensuremath{\Xires^0}}\xspace}

\def\Omegares    {{\ensuremath{\POmega}}\xspace}

\def\Lc          {{\ensuremath{\Lz^+_\cquark}}\xspace}

\def\Xic         {{\ensuremath{\Xires_\cquark}}\xspace}
\def\Xicz        {{\ensuremath{\Xires^0_\cquark}}\xspace}
\def\Xicp        {{\ensuremath{\Xires^+_\cquark}}\xspace}

\def\Lb           {{\ensuremath{\Lz^0_\bquark}}\xspace}

\def\Xib          {{\ensuremath{\Xires_\bquark}}\xspace}
\def\Xibz         {{\ensuremath{\Xires^0_\bquark}}\xspace}
\def\Xibm         {{\ensuremath{\Xires^-_\bquark}}\xspace}

\newcommand{\decay}[2]{\ensuremath{\mathinner{#1\!\to #2}}\xspace}

\def\to                 {\ensuremath{\rightarrow}\xspace}

\def\CP                {{\ensuremath{C\!P}}\xspace}

\def\AT#1     {\ensuremath{A_{\mathrm{T}}^{#1}}\xspace}           

\def\C#1      {\ensuremath{\mathcal{C}_{#1}}\xspace}                       
\def\Cp#1     {\ensuremath{\mathcal{C}_{#1}^{'}}\xspace}                    
\def\Ceff#1   {\ensuremath{\mathcal{C}_{#1}^{\mathrm{(eff)}}}\xspace}        
\def\Cpeff#1  {\ensuremath{\mathcal{C}_{#1}^{'\mathrm{(eff)}}}\xspace}       
\def\Ope#1    {\ensuremath{\mathcal{O}_{#1}}\xspace}                       
\def\Opep#1   {\ensuremath{\mathcal{O}_{#1}^{'}}\xspace}                    

\newcommand{\aunit}[1]{\ensuremath{\text{\,#1}}}       

\newcommand{\tev}{\aunit{Te\kern -0.1em V}\xspace}
\newcommand{\gev}{\aunit{Ge\kern -0.1em V}\xspace}
\newcommand{\mev}{\aunit{Me\kern -0.1em V}\xspace}
\newcommand{\kev}{\aunit{ke\kern -0.1em V}\xspace}
\newcommand{\ev}{\aunit{e\kern -0.1em V}\xspace}
 
\newcommand{\mevc}{\ensuremath{\aunit{Me\kern -0.1em V\!/}c}\xspace}
\newcommand{\gevc}{\ensuremath{\aunit{Ge\kern -0.1em V\!/}c}\xspace}
\newcommand{\mevcc}{\ensuremath{\aunit{Me\kern -0.1em V\!/}c^2}\xspace}
\newcommand{\gevcc}{\ensuremath{\aunit{Ge\kern -0.1em V\!/}c^2}\xspace}

\def\fb   {\ensuremath{\aunit{fb}}\xspace}
\def\invfb   {\ensuremath{\fb^{-1}}\xspace}

\newcommand{\stat}{\aunit{(stat)}\xspace}
\newcommand{\syst}{\aunit{(syst)}\xspace}

\def\gsim{{~\raise.15em\hbox{$>$}\kern-.85em
          \lower.35em\hbox{$\sim$}~}\xspace}
\def\lsim{{~\raise.15em\hbox{$<$}\kern-.85em
          \lower.35em\hbox{$\sim$}~}\xspace}

\def\sPlot{\mbox{\em sPlot}\xspace}

\def\tell1  {TELL1\xspace}
\def\ukl1   {UKL1\xspace}

\newcommand{\eg}{\mbox{\itshape e.g.}\xspace}

\newcommand{\phz}{\phantom{0}}

\newcommand{\lhcborcid}[1]{\href{https://orcid.org/#1}{\hspace*{0.1em}\raisebox{-0.45ex}{\includegraphics[width=1em]{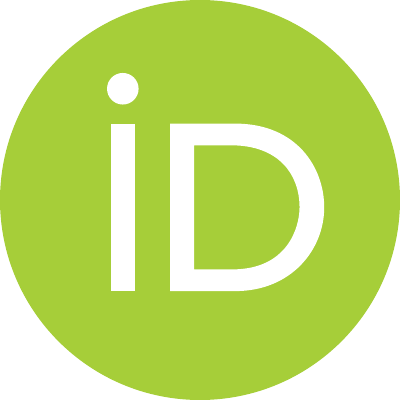}}}}

\hypersetup{
  colorlinks   = true, 
  urlcolor     = blue, 
  linkcolor    = blue, 
  citecolor    = red   
}

\ifEnableSectionTOCLinks
    \usepackage[explicit]{titlesec} 
    
    \let\oldcontentsline\contentsline
    \renewcommand

    \titleformat{\section}{\normalfont\Large\bf}{\hyperlink{tocsection.\thesection}{{\thesection} \parbox[t]{\dimexpr\textwidth-1pc}{#1}}}{1pc}{}

    \titleformat{\subsection}{\normalfont\bf}{\hyperlink{tocsubsection.\thesubsection}{{\thesubsection} \parbox[t]{\dimexpr\textwidth-1pc}{#1}}}{1pc}{}

    \titleformat{name=\section,numberless}[display]{}{}{0pt}{\normalfont\Huge\bfseries #1}
\fi

\usepackage{cite} 
\usepackage{LHCb/mciteplus}
\makeatletter
\g@addto@macro\bfseries{\boldmath}
\makeatother

\usepackage{longtable}

\begin{document}

\renewcommand{\thefootnote}{\fnsymbol{footnote}}
\setcounter{footnote}{1}

\begin{titlepage}
\pagenumbering{roman}

\vspace*{-1.5cm}
\centerline{\large EUROPEAN ORGANIZATION FOR NUCLEAR RESEARCH (CERN)}
\vspace*{1.5cm}
\noindent
\begin{tabular*}{\linewidth}{lc@{\extracolsep{\fill}}r@{\extracolsep{0pt}}}
\ifthenelse{\boolean{pdflatex}}
{\vspace*{-1.5cm}\mbox{\!\!\!\includegraphics[width=.14\textwidth]{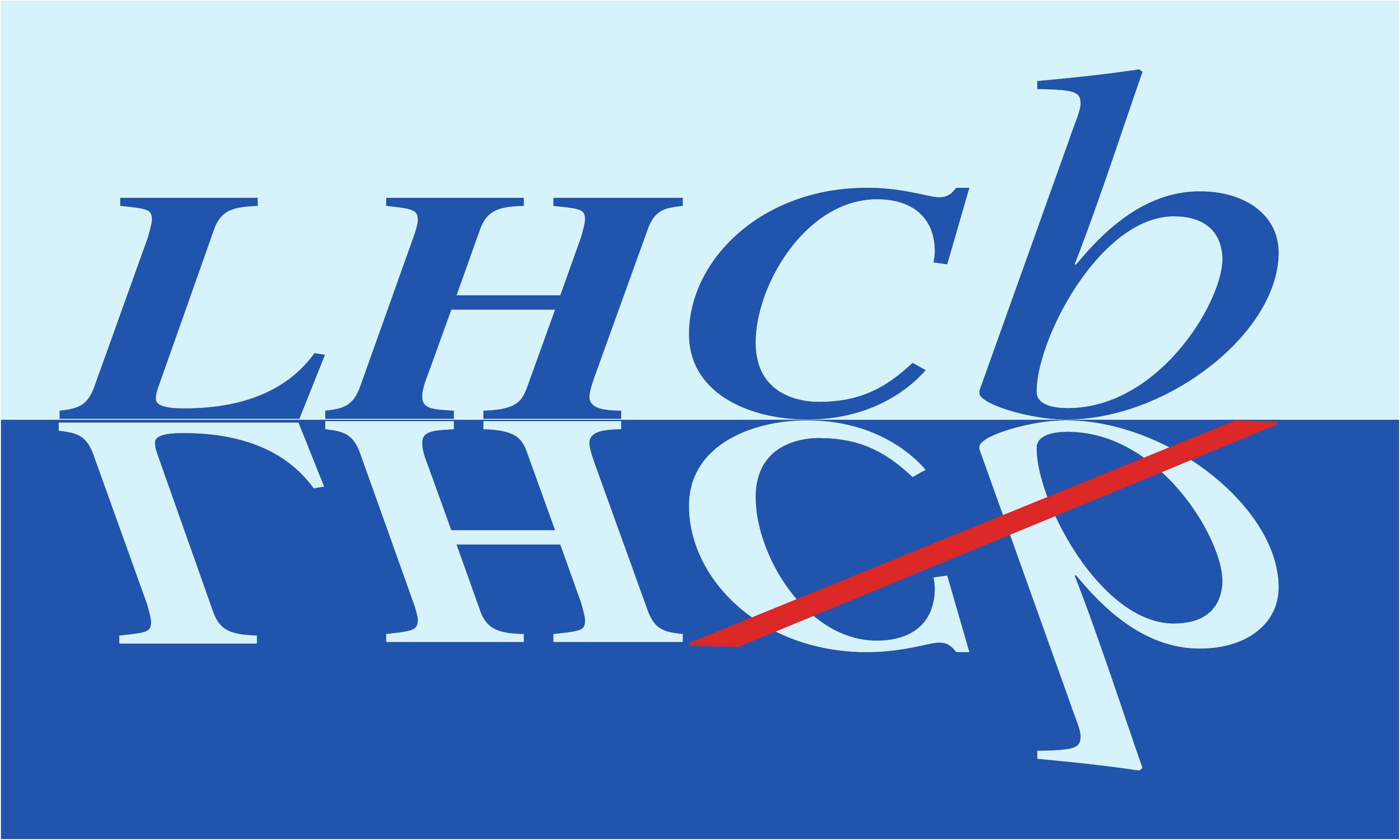}} & &}%
{\vspace*{-1.2cm}\mbox{\!\!\!\includegraphics[width=.12\textwidth]{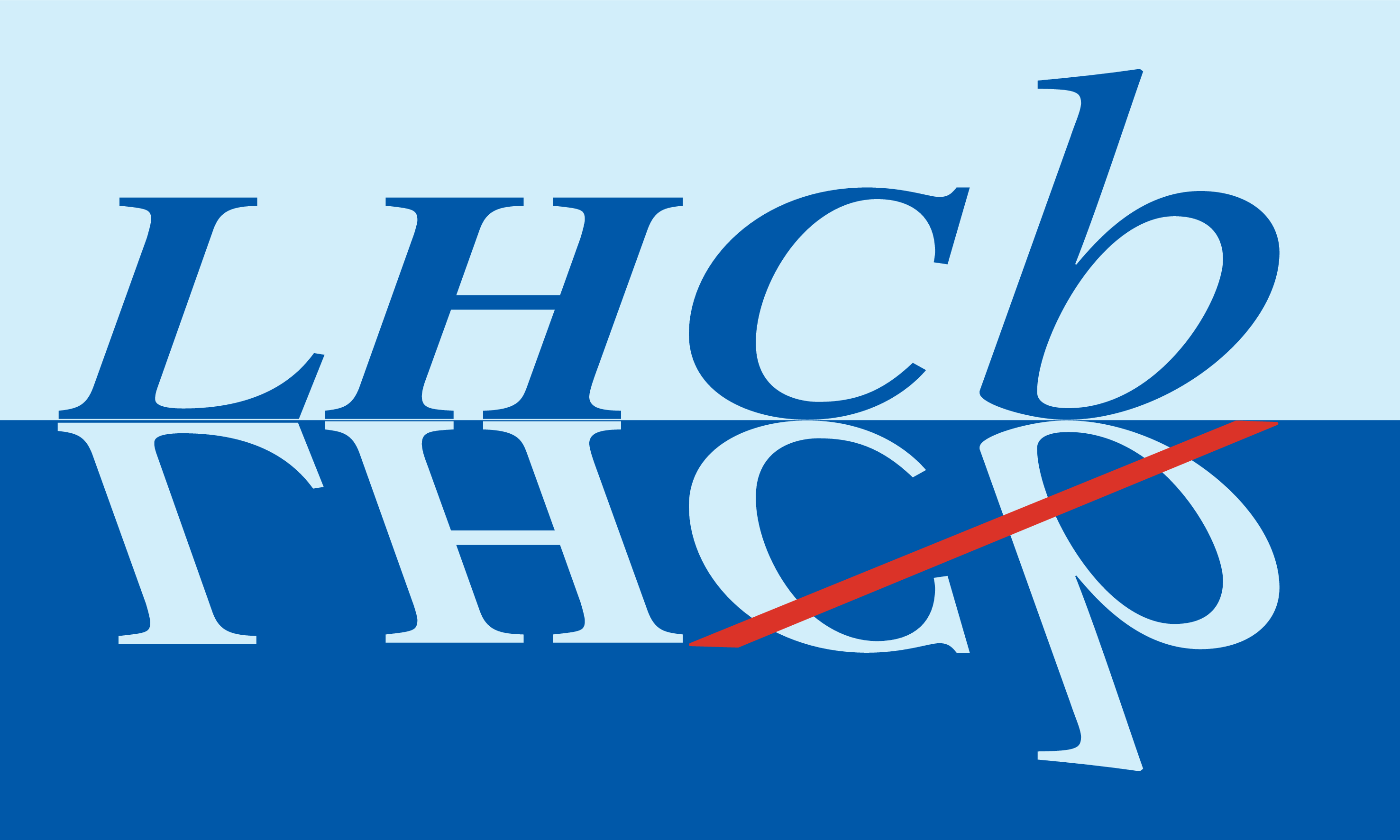}} & &}%
\\
 & & CERN-EP-2026-235 \\  
 & & LHCb-PAPER-2026-024 \\  
 & & August 27, 2026 \\ 
 & & \\
\end{tabular*}

\vspace*{4.0cm}

{\normalfont\bfseries\boldmath\huge
\begin{center}
  \papertitle 
\end{center}
}

\vspace*{2.0cm}

\begin{center}
\paperauthors\footnote{Authors are listed at the end of this Letter.}
\end{center}

\vspace{\fill}

\begin{abstract}
  \noindent   
A search for the Cabibbo-suppressed decay \decay{\Xicz}{p \Km} is performed using $pp$ collision data corresponding to an integrated luminosity of 5.4\invfb, collected by the \lhcb experiment at a centre-of-mass energy of 13\tev. The decay is observed for the first time and its branching fraction measured to be \mbox{$(4.5\pm0.5\pm0.2\pm0.9)\times10^{-5}$}, where the uncertainties are statistical, systematic and from the branching fraction of the normalisation channel \decay{\Xibm}{\Xicz (\decay{}{p \Km\Km \pip}) \pim}. Using the decay chain \decay{\Xibm}{\Xicz(\decay{}{p\Km})\pim}, the decay asymmetry parameter of the \decay{\Xicz}{p\Km} decay is determined to be $\alpha_{\Xicz}=0.32\pm0.15\pm0.01$.  
\end{abstract}

\vspace*{2.0cm}

\begin{center}
  Submitted to
  Phys.~Rev.~Lett. 
\end{center}

\vspace{\fill}

{\footnotesize 
\centerline{\copyright~\papercopyright. \href{\paperlicenceurl}{\paperlicence}.}}
\vspace*{2mm}

\end{titlepage}

\newpage
\setcounter{page}{2}
\mbox{~}


\renewcommand{\thefootnote}{\arabic{footnote}}
\setcounter{footnote}{0}

\cleardoublepage

\pagestyle{plain} 
\setcounter{page}{1}
\pagenumbering{arabic}


Charm-baryon decays are invaluable for studying the interplay between low-energy strong and weak interactions within the Standard Model. They are sensitive to key dynamical features such as colour suppression, internal and external $W$-boson emission, exchange, and final-state interaction processes~\cite{Cheng:2021qpd,Cheng:2024lsn}. Purely hadronic weak decays are of particular interest as they enable tests of the approximate SU(3) flavour symmetry and investigations of nonfactorisable contributions to the decay amplitudes at the charm-quark mass scale~\cite{Niu:2020gjw}. The decay asymmetry parameters are also relevant for potential polarization measurements with these channels, providing sensitivity to charm-quark production mechanisms and hadronisation in hadron collisions~\cite{LHCb-PAPER-2024-017}. Precise experimental inputs are particularly important for singly Cabibbo-suppressed (SCS) modes, where theoretical predictions are limited by nonperturbative QCD effects~\cite{Niu:2020gjw}.

Motivated by this need, important experimental progress has been achieved on the properties of charm-baryon decays, including both branching fractions and decay asymmetry parameters. Several SCS \Xicp and $\Omegares_c^0$ decays have been observed by the \lhcb, \belle and \belletwo experiments\cite{LHCb-PAPER-2023-011,Belle:2024xcs}, providing new insights into charm-baryon decays. Charge-conjugated states and decays are implied throughout this Letter unless otherwise stated. 

Originally introduced to probe parity violation in hyperons~\cite{Lee:1957qs},  decay asymmetry parameters offer sensitivity to parity violation and strong-interaction dynamics that cannot be inferred from branching fractions alone~\cite{Zhang:2026qjt}.
Recently, decay asymmetry parameters have been determined for several charm-baryon decays~\cite{LHCb-PAPER-2024-017,BESIII:2023wrw,BESIII:2019odb,Belle:2022uod,Tang:2022ksv}, giving direct access to the relative magnitudes and phases of decay amplitudes.

\begin{figure}[hbp]
    \centering
    \captionsetup{justification=raggedright,singlelinecheck=false}
    \begin{subfigure}{0.36\textwidth}
        \centering
        \includegraphics[
            width=\textwidth,
            trim={10 2 5 3},  
            clip
        ]{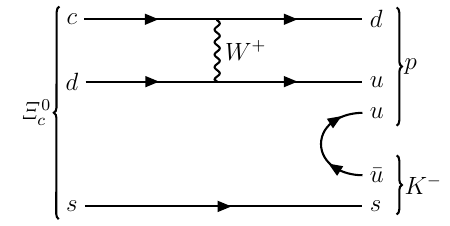}
        \label{fig:tree}
    \end{subfigure}
    \hspace{0.06\textwidth}
    \begin{subfigure}{0.36\textwidth}
        \centering
        \includegraphics[
            width=\textwidth,
            trim={2 4 1 0},  
            clip
        ]{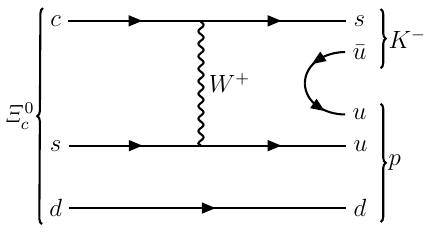}
        \label{fig:tree2}
    \end{subfigure}
    \caption{Leading Feynman diagrams contributing to the $\Xicz \to \proton \Km$ decay, representing the $W$-exchange contributions of this Cabibbo-suppressed process.}
    \label{fig:fey_dia}
\end{figure}

Despite this progress, a precise theoretical description of charm-baryon decays remains elusive due to nonperturbative effects and the limited set of measurements, in particular for nonfactorisable $W$-exchange decays, which cannot be computed reliably. 
Notably, purely $W$-exchange SCS decays of the \Xic baryon, \eg\ the \(\Xicz \to p\Km\) decay of  Fig.~\ref{fig:fey_dia}, have never been observed. 
The decay asymmetry parameter, defined as $\alpha_{\Xicz} \equiv 2\,\mathrm{Re}(S^{*}P)/(|S|^{2}+|P|^{2})$, where $S$ and $P$ denote the $S$- and $P$-wave amplitudes of the \(\Xicz \to p\Km\) decay and $S^*$ denotes the complex conjugate of $S$, quantifies the interference between the parity-violating $S$-wave and parity-conserving $P$-wave amplitudes~\cite[Sec.~79]{PDG2024}.  This parameter  provides both insight into the underlying hadronic dynamics and sensitivity to potential \CP violation in charm baryons~\cite{Yang:2025orn}.
Moreover, the recent full angular analysis of the related $W$-exchange-dominated two-body decay \decay{\Lc}{\Xiz\Kp} by \besiii~\cite{BESIII:2023wrw} shows deviations from theoretical predictions, suggesting the limited reliability of current models. A measurement of both the branching fraction and the decay asymmetry parameter of \decay{\Xicz}{\proton\Km} decay is therefore essential for testing these predictions and improving our understanding of charm-baryon decay dynamics.

This Letter reports the first observation of the SCS decay \decay{\Xicz}{\proton\Km}, using \Xicz baryons produced in the \decay{\Xibm}{\Xicz\pim} decay. The cascade decay provides a lower-background sample and uniquely enables the measurement of the \Xicz decay asymmetry parameter. The analysis is performed using proton-proton ($pp$) collision data collected at a centre-of-mass energy of 13\tev, corresponding to an integrated luminosity of 5.4\invfb.

The \decay{\Xicz}{\proton \Km} branching fraction is determined relative to that of the normalisation channel \decay{\Xicz}{\proton\Km\Km\pip}, and the decay asymmetry parameter is measured through an angular analysis of the sequential \decay{\Xibm}{\Xicz(\decay{}{\proton \Km})\pim} decay.

The \lhcb detector is a single-arm forward spectrometer covering the pseudorapidity range $2 < \eta < 5$, as described in detail in Refs.~\cite{LHCb-DP-2008-001,LHCb-DP-2014-002}.
The online selection of $\Xibm$ decays is performed by the LHCb trigger system~\cite{LHCb-DP-2012-004}, which consists of a hardware stage followed by a software stage~\cite{LHCb-TDR-017,LHCb-PROC-2015-018,BBDT,LHCb-DP-2019-001}. The hardware trigger is based on information from the calorimeter and muon systems. The software trigger selects events containing a secondary vertex formed by two or three tracks significantly displaced from any primary $pp$ collision vertices (PVs).

Simulated samples of \decay{\Xibm}{\Xicz(\decay{}{ p\Km})\pim} and \decay{\Xibm}{\Xicz(\decay{}{ p\Km\Km\pip})\pim} decays are used to optimise the offline selection, study potential backgrounds and model the detector acceptance. 
The simulated samples are produced using the software described in Refs.~\cite{Sjostrand:2007gs,LHCb-PROC-2010-056,Lange:2001uf,davidson2015photos,Allison:2006ve,LHCb-PROC-2011-006}.
The decay products in the \Xibm cascade are generated uniformly over the allowed phase space. 
The simulated kinematic spectra of the \Xibm baryons,  angular distributions of the final-state particles and  event track multiplicity are corrected to reproduce those observed in data. 
The particle-identification (PID) response is further calibrated using dedicated data samples~\cite{LHCb-DP-2018-001}.

In the offline selection, all final-state tracks are required to have transverse momentum larger than $200\mevc$ and to be inconsistent with originating from every PV.
The  \Xicz candidates reconstructed in either the \decay{\Xicz}{p\Km} or \decay{\Xicz}{p\Km\Km\pip} decays are required to have a well-fitted secondary vertex significantly displaced from all PVs.
The \Xicz invariant mass is required to be within $\pm20\mevcc$ ($\pm15\mevcc$), corresponding to about three times the mass resolution obtained from simulation, of the known \Xicz mass~\cite{PDG2024} for the \decay{\Xicz}{\proton \Km} (\decay{\Xicz}{\proton \Km\Km\pip}) decay, where the tighter window reflects the better mass resolution of the four-body decay due to its smaller phase space. The \Xicz candidate is then combined with a track identified as a pion to form the \Xibm candidate.
The \Xibm invariant mass is obtained using a kinematic fit~\cite{Hulsbergen:2005pu} that constrains the \Xicz mass to its known value and requires the reconstructed \Xibm momentum to point to its best-matched PV.

The \Bm meson decays to final states such as \Kp\Km\Km, \Kp\Km\pim and \pip\Km\pim can be falsely reconstructed as \decay{\Xibm}{\Xicz(\decay{}{\proton \Km})\pim} if kaons or pions are misidentified. These contributions are removed using tight PID requirements and corresponding mass spectrum vetoes.
For the \decay{\Xicz}{\proton\Km} channel, additional backgrounds arising from \Dz meson decays to \Kp\Km, \Kp\pim and \pip\Km final states are removed using vetoes on the corresponding invariant-mass spectra. For the \decay{\Xicz}{\proton \Km\Km\pip} decay, backgrounds from \decay{\phi(1020)}{\Kp\Km} decays, where one kaon is misidentified as a proton, and from \decay{\Dz}{\Kp\Km\Km\pip} decays are removed similarly.

A multivariate classifier based on a boosted decision tree (BDT)~\cite{Breiman,AdaBoost} is used to separate signal from combinatorial background for each $\Xibm$ decay mode. 
The BDT is trained using simulated signal samples and background candidates selected from the high-mass sideband from $6000$ to $6600\mevcc$ of the $\Xibm$ invariant-mass spectrum. A requirement on $m(\Xicz\pim)$ in the range $[5600,6000]\mevcc$ is applied after the BDT selection. 
The BDT classifier uses variables describing the kinematics and topology of the \Xibm decay chain, together with PID information of the final-state particles.
The BDT requirements for both \decay{\Xibm}{ \Xicz(\decay{}{\proton \Km})\pim} and \decay{\Xibm}{ \Xicz(\decay{}{\proton \Km\Km\pip})\pim} decays are optimised by maximising the figure of merit $N_S/\sqrt{N_S+N_B}$. The variables $N_S$ and $N_B$  represent the signal and  background yields in the \Xibm signal region, defined as  $\pm 3\sigma$ around the known \Xibm mass~\cite{PDG2024}, where $\sigma$ is the invariant-mass resolution, approximately 14\mevcc.

The yields of the \decay{\Xibm}{\Xicz\pim} decays, with \decay{\Xicz}{\proton \Km} and \decay{\Xicz}{ \proton \Km\Km\pip}, are determined from a simultaneous unbinned maximum-likelihood fit to the two \Xibm invariant mass spectra. 
Four contributions are considered in the spectra, namely signal and the combinatorial, misidentified and partially reconstructed backgrounds. 
The signal is modelled as the sum of a double-sided Crystal Ball (DSCB) function~\cite{Skwarnicki:1986xj} and a Gaussian function, sharing a common mean but with different widths. 
The DSCB tail parameters, the relative fraction of the two subcomponents and the ratio of their widths are fixed to values obtained from simulated decays. 
The signal models for the two separate \Xicz decays share a common peak position. The ratio of their invariant-mass resolutions is fixed to the value in simulation.
The combinatorial background is described by an exponential function with an independent slope parameter in each decay mode. 
The misidentification background from the \decay{\Xibm}{\Xicz\Km} decay is described by the sum of a DSCB function and a Gaussian function, with parameters fixed from simulation.
The partially reconstructed background due to the \decay{\Xibm}{\Xicz\pip\piz} decay is modelled using an ARGUS function~\cite{ARGUS:1990hfq} convolved with a Gaussian resolution function. 
The ratio of the yields of the misidentification and the signal is shared between the two \Xicz samples, as is the ratio of the yields of the partially reconstructed background and the signal.
 
\begin{figure}[htbp]
        \centering
\includegraphics[width=0.48\textwidth]{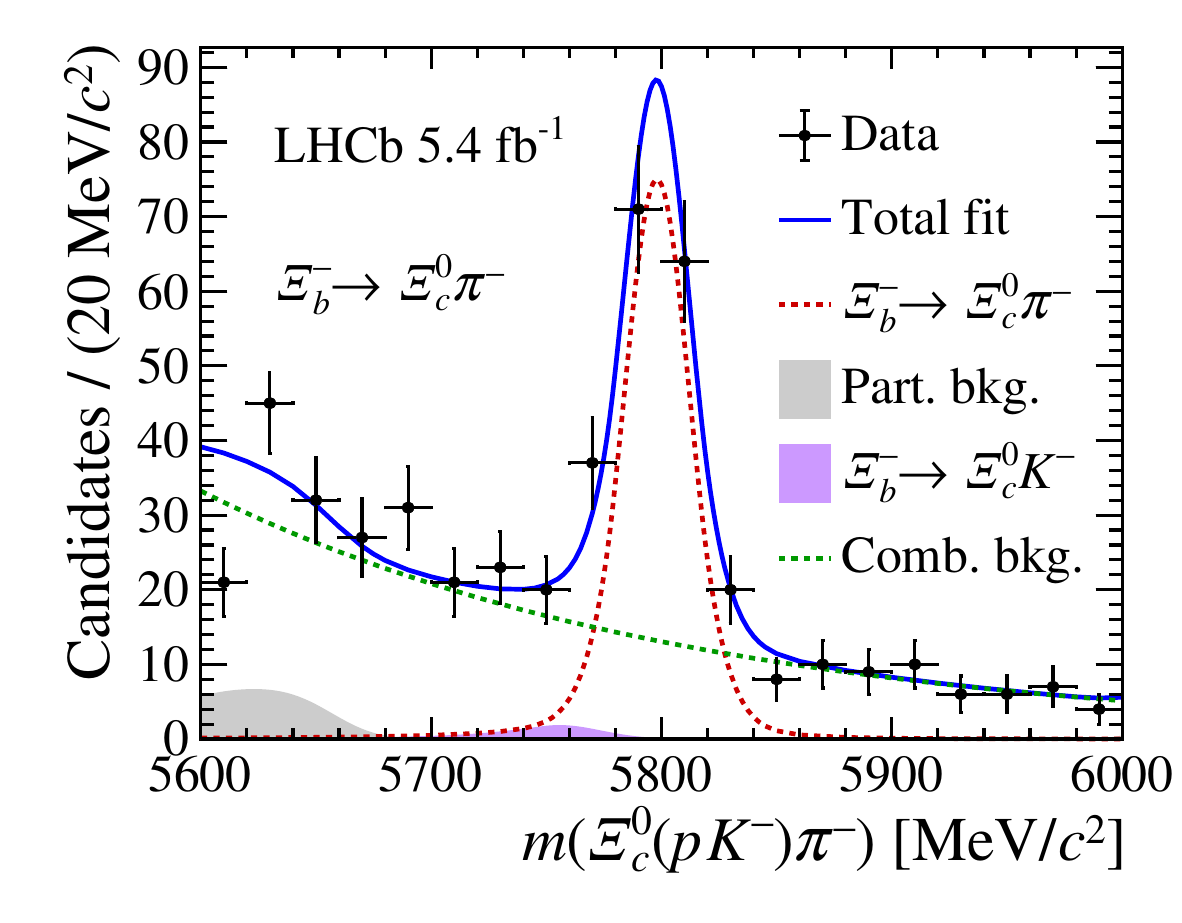}
\includegraphics[width=0.48\textwidth]{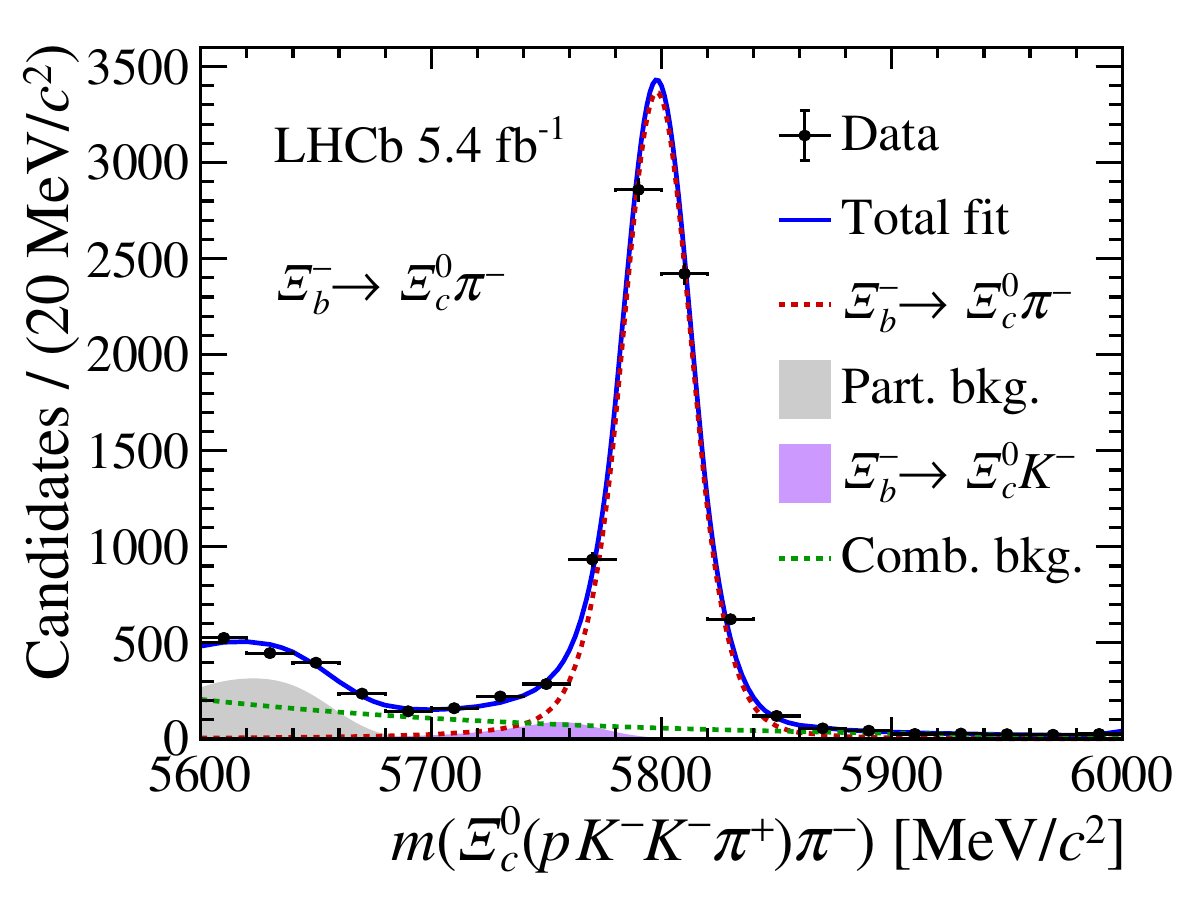}
\includegraphics[width=0.48\textwidth]{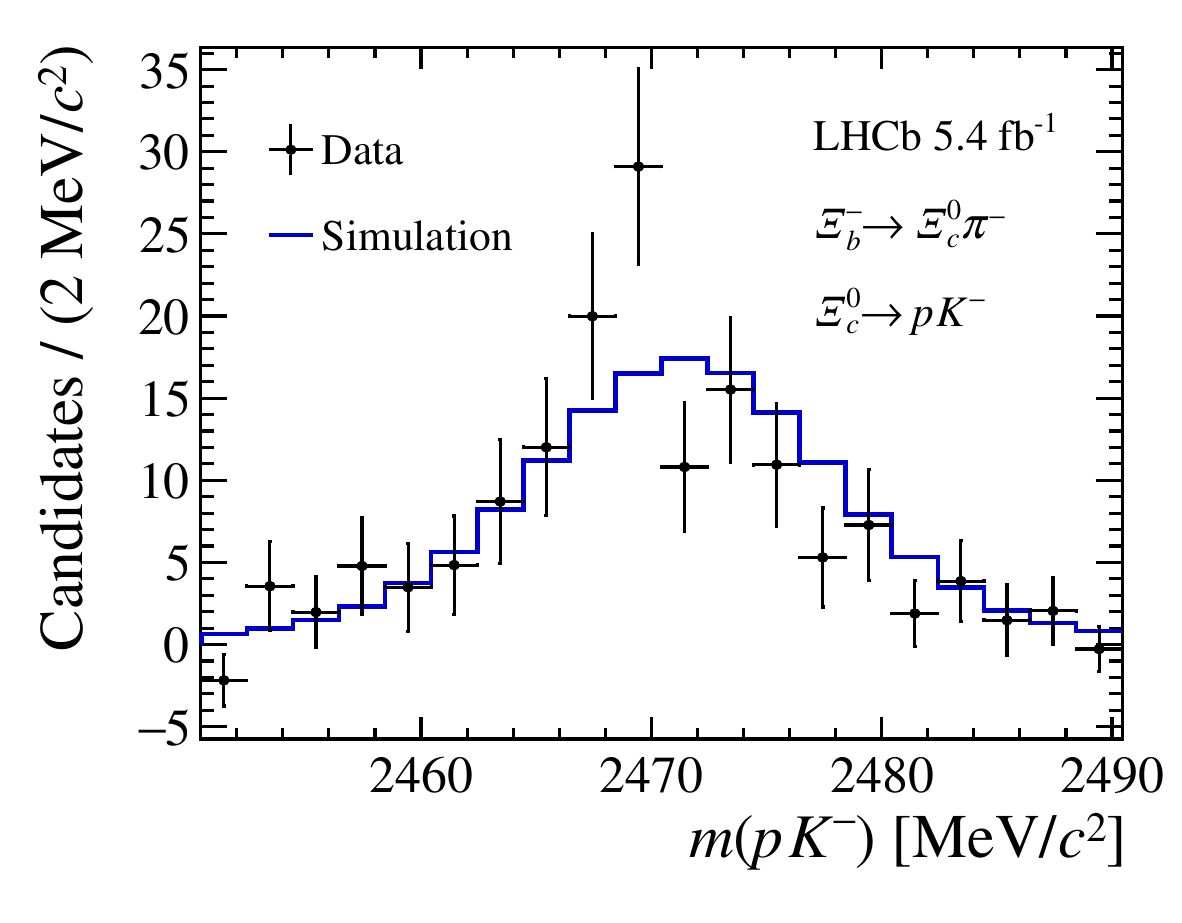}
\includegraphics[width=0.48\textwidth]{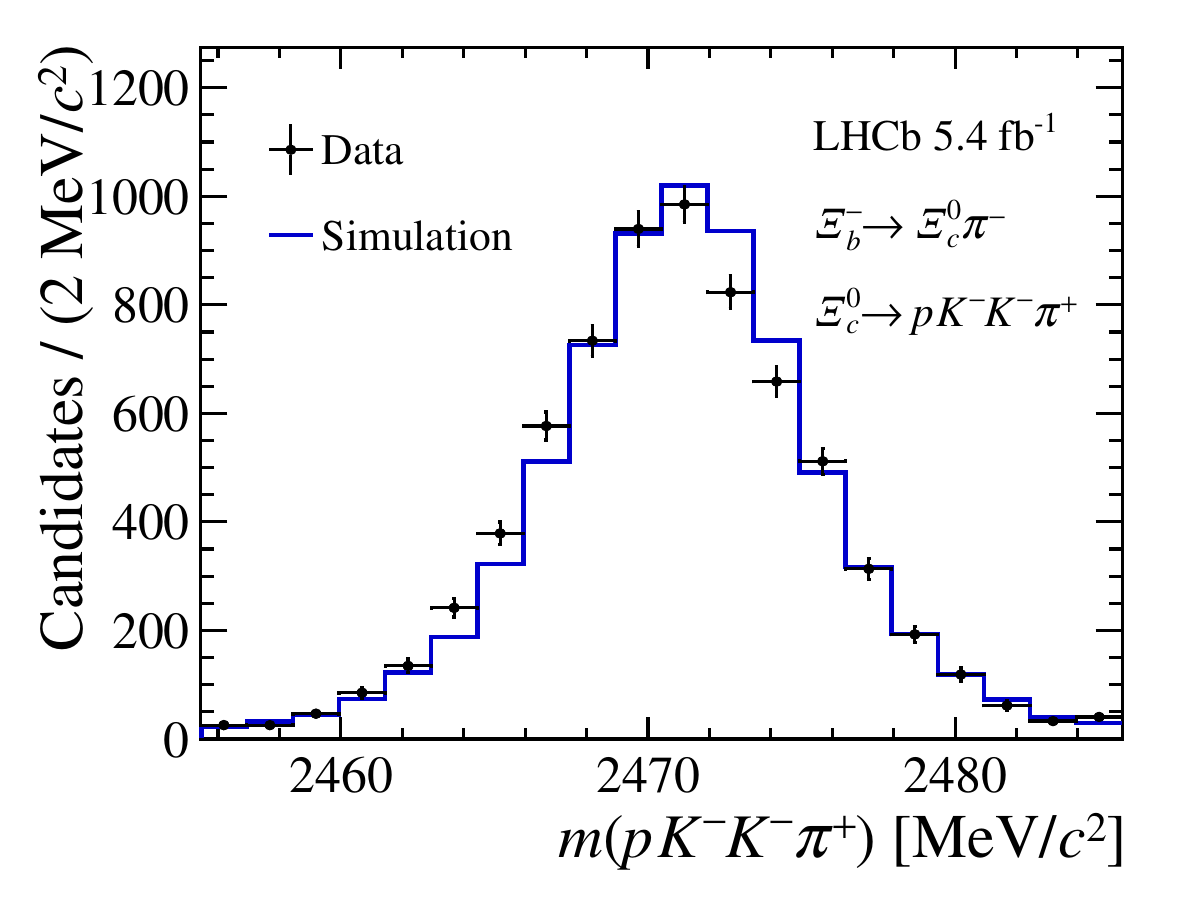} 
    \caption{Invariant-mass distributions of (top) $\Xibm$ and (bottom) background-subtracted \Xicz  candidates reconstructed in the (left) \decay{\Xibm}{\Xicz(\decay{}{\proton \Km})\pim} and  (right) \decay{\Xibm}{\Xicz(\decay{}{\proton \Km\Km\pip})\pim}) decay modes. The corresponding simulated distributions after kinematic corrections are overlaid.}
    \label{fig:mass_fit}
\end{figure}

Figure~\ref{fig:mass_fit} shows the invariant-mass distributions of the selected candidates. The signal yields for the \decay{\Xibm}{\Xicz(\decay{}{\proton \Km})\pim} and \decay{\Xibm}{\Xicz(\decay{}{\proton \Km\Km\pip})\pim} decays are $134 \pm 14$ and $(6.93\pm0.09)\times10^{3}$, respectively, where the uncertainties are statistical. This constitutes the first observation of the \decay{\Xicz}{\proton\Km} decay.

The background-subtracted mass distributions for the \proton\Km signal and \proton\Km\Km\pip normalisation channels are shown in Fig.~\ref{fig:mass_fit}. The distributions are consistent in shape with the simulated \Xicz signal in the signal and normalisation channels, respectively, supporting a resonant \Xicz interpretation of the observed \Xibm signal.

The branching fraction of the \decay{\Xicz}{ \proton \Km} decay is measured relative to that of the \decay{\Xicz}{\proton \Km\Km\pip} decay using 
\begin{equation}
    \mathcal{R} \equiv \frac{\mathcal{B}(\Xicz\to \proton \Km)}{\mathcal{B}(\Xicz\to \proton \Km\Km\pip)}  
    = \frac{N_{\text{sig}}}{\epsilon_{\text{sig}}} / \frac{N_{\text{norm}}}{\epsilon_{\text{norm}}},
    \label{eq:branch_fraction}
\end{equation}
where `sig' and `norm' represent the \decay{\Xibm}{\Xicz(\decay{}{\proton \Km})\pim} and  \decay{\Xibm}{\Xicz(\decay{}{\proton \Km\Km\pip})\pim} decays, respectively.
Here $N$ denotes the yield of the signal or normalisation channel, extracted from the data fits, and $\epsilon$ represents the total efficiency, including the trigger, reconstruction and selection, determined from the corrected simulation samples.
The \decay{\Xibz}{\Xicz\pim} branching fraction and the \Xibz production cross-section cancel in this expression. The relative branching fraction is measured to be $\mathcal{R} = (9.26\pm0.94)\times 10^{-3}$. The fit results are used to compute the signal weight for each candidate via the \sPlot technique~\cite{Pivk:2004ty}, which is applied to subtract background contributions in the subsequent angular analysis. The result of the angular analysis is used to correct the simulated angular distribution, with its uncertainty propagated as an associated systematic uncertainty in the branching fraction measurement.

The decay asymmetry parameter of the \decay{\Xicz}{\proton \Km} decay is encoded in the distribution of the $\Xicz$ helicity angle, $\theta$, defined as the angle between the proton momentum in the \Xicz rest frame and the \Xicz momentum in the \Xibm rest frame~\cite{Jacob:1959at}.
Previous LHC measurements~\cite{LHCb-PAPER-2012-057,CMS:2018wjk,LHCb-PAPER-2020-005} indicate that the \bquark-baryons produced in $pp$ collisions at LHC energies are consistent with being unpolarised, which is verified through a study of the normalisation channel in this analysis. 
Thus the $\theta$ distribution can be parametrised as
\begin{equation}
\label{equ:ang_dis}
\frac{\mathrm{d}N}{\mathrm{d}\cos\theta}
\propto 1 + \alpha_{\Xib}\alpha_{\vphantom{\Xib}\Xicz}\cos\theta,
\end{equation}
where \(\alpha_{\Xib}\) is the decay asymmetry parameter of the \decay{\Xibm}{\Xicz\pim} decay, 
predicted to have a value of $-1$~\cite{Cheng:1996cs,Li:2021kfb}.
A recent measurement for the \decay{\Lb}{\Lc\pim} decay~\cite{LHCb-PAPER-2024-017}, dominated by dynamics similar to that of the \decay{\Xibm}{\Xicz\pim} decay, validates the prediction. Therefore $\alpha_{\Xib} = -1$ is assumed hereafter.

\begin{figure}[tb] 
    \centering
    \includegraphics[width=0.65\columnwidth]{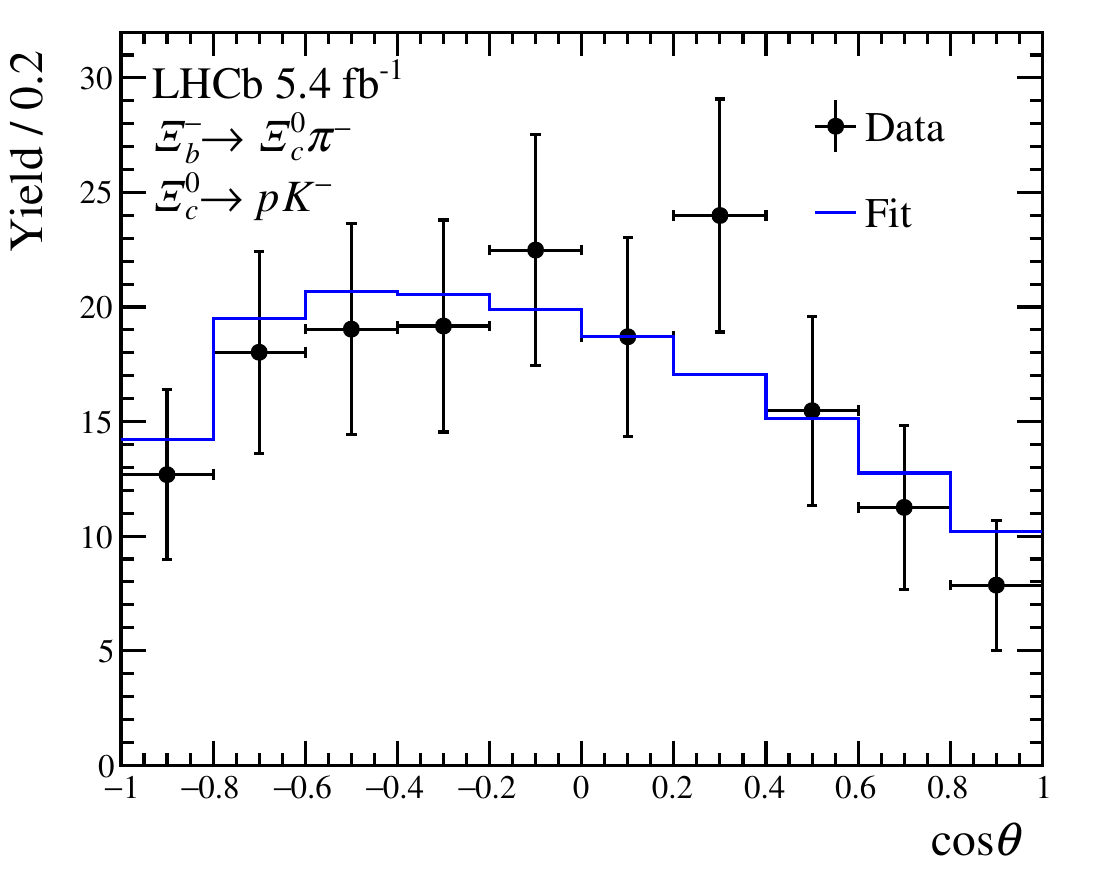}
    \caption{Background-subtracted distribution of the cosine of the $\Xicz$ helicity angle for the \decay{\Xibm}{\Xicz(\decay{}{\proton \Km})\pim} decay. The curve represents the fit result from Eq.~\eqref{equ:ang_dis}.}    
    \label{fig:ang_fit}
\end{figure}

The logarithm of the likelihood function for the angular fit is defined using the weighted likelihood formalism~\cite{Xie:2009rka} as  
\begin{equation}
    \ln\mathcal{L}( \alpha_{\Xib}\alpha_{\vphantom{\Xib}\Xicz})
    = 
    \frac{\sum_i w_i}{\sum_i w_i^2}\,
    \sum_{i=1}^{N}
    w_i \,
    \ln\!\left[\mathcal{P}(\cos\theta_i \,|\,  \alpha_{\Xib}\alpha_{\vphantom{\Xib}\Xicz})\right],
\label{equ:paper_ll}
\end{equation}
where $w_i$ are the \sPlot weights used to statistically subtract background contributions and  
the sum extends over the data sample of \decay{\Xibm}{ \Xicz(\decay{}{p\Km})\pim} candidates. 
The scale factor \(\sum_i w_i/\sum_i w_i^2\) is used to correct the statistical uncertainty for the weighted likelihood fit~\cite{Langenbruch:2019nwe}.
The probability density function (PDF), $\mathcal{P}(\cos\theta| \alpha_{\Xib}\alpha_{\vphantom{\Xib}\Xicz})$, is the product of the angular distribution from Eq.~\eqref{equ:ang_dis} and the experimental efficiency $\epsilon(\cos\theta)$, which accounts for the angular dependence induced by the detector acceptance and selections. The efficiency is determined from simulated decays as a function of $\cos\theta$ and used in the normalisation of the PDF following the procedure in Ref.~\cite{LHCb-PAPER-2013-008}.
Figure~\ref{fig:ang_fit} shows the angular distribution of the \decay{\Xibm}{\Xicz(\decay{}{ p\Km})\pim} decay and the projection of the fit, yielding $\alpha_{\Xicz}=0.32\pm0.15\stat$.

The main systematic uncertainties in the relative branching fraction ratio and decay asymmetry parameter are summarised in Table~\ref{table:Rsys_uncertainties}. 
The systematic uncertainty from the modelling of the signal and background mass shapes is evaluated using alternative models. 
Additional systematic uncertainties are estimated from PID efficiency, trigger efficiency and kinematic corrections. 
These contributions are evaluated by varying the corresponding corrections within their uncertainties. 
The uncertainty on the efficiency due to the limited size of the simulated samples is propagated to the measurements by repeatedly resampling the simulated samples.
The uncertainty from charmless background accounts for possible contributions from direct \Xibm decays to the same final state without an intermediate \Xicz resonance. 
This contribution is estimated using data in the \Xicz sideband regions, which are well separated from the \Xicz signal window.
The uncertainty from possible \Xibm polarisation on the decay asymmetry parameter is evaluated by assigning a 5\% polarisation for the \Xibm baryon, which is taken as a conservative estimate based on previous LHCb measurements of \Lb baryons~\cite{LHCb-PAPER-2020-005}.
For the relative branching fraction, the different number of charged tracks in the signal and normalisation channels introduces an uncertainty due to imperfect modelling of the tracking efficiency, estimated to be 1.5\% per track. 
In addition, the uncertainty on the proton helicity angle distribution, driven by the decay asymmetry parameter, is a source of systematic uncertainty for the relative branching fraction measurement due to its nonuniform efficiency, which is evaluated by varying the parameter within one standard deviation.
The total systematic uncertainty is the sum in quadrature of all sources considered, determined to be $0.01$ for the decay asymmetry parameter and $0.43\times10^{-3}$ for the branching fraction measurement, giving
\begin{align*}
    \alpha_{\Xicz}&=\phantom{(}0.32\pm0.15 \stat \pm0.01\syst,\\
    \mathcal{R} &= (9.26\pm0.94\stat\pm0.43 \syst)\times 10^{-3}.
\end{align*}
The measured value of \(\alpha_{\Xicz}\) is compatible with zero within the uncertainties.
Given the known value of $\mathcal{B}( \Xicz \to p \Km \Km \pip)=(4.9\pm1.0)\times10^{-3}$~\cite{PDG2024}, the $\Xicz \to p \Km$ branching fraction is determined to be 
\begin{equation*}
\mathcal{B}( \Xicz \to p \Km )=(4.5\pm0.5 \stat\pm0.2\syst  \pm0.9\,\text{(norm)})\times10^{-5}, 
\end{equation*}
where the third uncertainty is from knowledge of the \decay{\Xicz}{\proton \Km \Km \pip} branching fraction.

\begin{table}[tb]
  \centering
  \caption{
  Systematic uncertainties in the relative branching fraction ratio and the decay asymmetry parameter measurement. The total uncertainties are the sum in quadrature of all contributions.}
  \label{table:Rsys_uncertainties}
  
  \begin{tabular}{@{}l c c@{}}
    \hline
    Source                            & $\mathcal{R}(\times10^{-3})$ &$\alpha_{\Xicz}$\\ 
    \hline
    Fit model                         & $0.20$  & $0.001$  \\
    PID efficiency                    & $0.05$  & $0.003$  \\
    Trigger efficiency                & $0.02$  & $0.001$  \\
    Kinematic correction              & $0.26$  & $0.001$  \\
    Simulation sample size            & $0.05$  & $0.003$  \\
    Charmless background              & $0.20$  & $0.007$  \\
    $\Xibm$ polarisation              & ---     & $0.003$  \\
    Tracking efficiency               & $0.19$  & --- \\
    Angular distribution              & $0.04$  & ---      \\
    \hline
    Total                             & $0.43$  & $0.009$  \\
    \hline
  \end{tabular}
\end{table}

The branching fraction and decay asymmetry parameter of the \mbox{$\Xicz \to p\Km$} decay, calculated using the irreducible SU(3) approach (IRA), the topological diagram approach (TDA) and current algebra (CA), are summarized in Table~\ref{tab:sum}.
The measured branching fraction is significantly lower than most theoretical predictions, except for the SU(3)$^{b}$ result whose uncertainty is too large to be constraining. The measured decay asymmetry parameter is compatible with IRA/TDA and SU(3)-based predictions in Ref.~\cite{Cheng:2024lsn} within about two standard deviations, but differs significantly from IRA and CA predictions in Refs.~\cite{Yang:2025orn,Zou:2019kzq}.

\begin{table}[tbhp]
	\centering
	\caption{Theoretical predictions of the branching fraction and $\alpha_{\Xicz}$ for \decay{\Xicz}{\proton \Km} decays. The superscript $a$ denotes a model with SU(3) flavour symmetry, while model $b$ includes SU(3) symmetry-breaking effects.}
	\label{tab:sum}
	\begin{tabular}{c c c}
	\hline
    Theory  & $\mathcal{B}(\Xicz\to p\Km)$  & $\alpha_{\Xicz}$\\
            & ($\times10^{-4}$) & \\
    \hline
    IRA~\cite{Cheng:2024lsn} & $3.1 \pm 0.2$ & $-0.07\pm0.11$ \\
    TDA~\cite{Cheng:2024lsn} & $3.1 \pm 0.2$ & $-0.07\pm0.10$ \\
    IRA~\cite{Yang:2025orn} & $2.7 \pm 0.3$ & $-0.81\pm0.11$ \\ 
    TDA~\cite{Zhao:2018mov} & $2.5 \pm 1.6$ & -- \\
    CA~\cite{Zou:2019kzq} & $3.5$ & $0.99$ \\
    SU(3)~\cite{Geng:2019xbo} & $5.0 \pm 1.1$ & $0.67\pm0.17$ \\
    SU(3)$^{a}$~\cite{Zhong:2022exp} & $1.38^{\:+\:0.35}_{\:-\:0.47}\phz$ & $0.65^{\:+\:0.34}_{\:-\:0.32}\phz$ \\[0.3ex]
    SU(3)$^{b}$~\cite{Zhong:2022exp} & $\,4^{\:+\:29}_{\:-\:23}$ & $0.086\pm0.031$ \\
    \hline
	\end{tabular}
\end{table}

In summary, a search is performed for the SCS decay \decay{\Xicz }{p\Km}, using \decay{\Xibm}{\Xicz\pim} decays selected in $pp$ collision data at a centre-of-mass energy of 13\tev, corresponding to an integrated luminosity of 5.4\invfb. A clear excess of signal candidates is observed with overwhelming significance, representing the first observation of this decay mode. 
The branching fraction and decay asymmetry parameter are measured for the \decay{\Xicz}{p\Km} decay.
The decay asymmetry parameter being consistent with zero, indicates no significant evidence for interference between the $S$- and $P$-wave amplitudes in this purely $W$-exchange channel, and thus places direct constraints on the underlying nonperturbative QCD dynamics.
The measurements show sizeable differences from most theoretical predictions, providing important input for improving theoretical descriptions of charm-baryon decays. 
With the substantially increased datasets collected by the upgraded LHCb detector, more precise measurements will become available to constrain theoretical models and may give access to the \CP violation in the decay asymmetry parameter.

\section*{Acknowledgements}
%
%
\noindent We express our gratitude to our colleagues in the CERN
accelerator departments for the excellent performance of the LHC. We
thank the technical and administrative staff at the LHCb
institutes.
We acknowledge support from CERN and from the national agencies:
ARC (Australia);
CAPES, CNPq, FAPERJ and FINEP (Brazil); 
MOST and NSFC (China); 
CNRS/IN2P3 and CEA (France);  
BMFTR, DFG and MPG (Germany);
NKFIH (Hungary);              
INFN (Italy); 
NWO (Netherlands); 
MNiSW and NCN (Poland); 
MEC/IFA (Romania); 
MICIU and AEI (Spain);
SNSF and SER (Switzerland); 
NASU (Ukraine); 
STFC (United Kingdom); 
DOE NP and NSF (USA).
We acknowledge the computing resources that are provided by ARDC (Australia), 
CBPF (Brazil),
CERN, 
IHEP and LZU (China),
IN2P3 (France), 
KIT and DESY (Germany), 
INFN (Italy), 
SURF (Netherlands),
Polish WLCG (Poland),
IFIN-HH (Romania), 
PIC (Spain), CSCS (Switzerland), 
GridPP (United Kingdom),
and NSF (USA).  
We are indebted to the communities behind the multiple open-source
software packages on which we depend.
Individual groups or members have received support from
RTP (Australia), 
FWO Odysseus grant G0ASD25N (Belgium), 
Key Research Program of Frontier Sciences of CAS, CAS PIFI, CAS CCEPP (China); 
Minciencias (Colombia);
EPLANET, Marie Sk\l{}odowska-Curie Actions, ERC and NextGenerationEU (European Union);
A*MIDEX, ANR, IPhU and Labex P2IO, and R\'{e}gion Auvergne-Rh\^{o}ne-Alpes (France);
Alexander-von-Humboldt Foundation (Germany);
ICSC (Italy); 
Severo Ochoa and Mar\'ia de Maeztu Units of Excellence, GVA, XuntaGal, GENCAT, InTalent-Inditex and Prog.~Atracci\'on Talento CM (Spain);
the Leverhulme Trust, the Royal Society and UKRI (United Kingdom).

\clearpage

\clearpage

\clearpage
\addcontentsline{toc}{section}{References}
\bibliographystyle{LHCb/LHCb}
\bibliography{main,LHCb/standard,LHCb/LHCb-PAPER,LHCb/LHCb-CONF,LHCb/LHCb-DP,LHCb/LHCb-TDR}

\newpage
\clearpage
\centerline
{\large\bf LHCb collaboration}
\begin
{flushleft}
\small
R.~Aaij$^{39}$\lhcborcid{0000-0003-0533-1952},
M.~Abdelfatah$^{71}$,
A.S.W.~Abdelmotteleb$^{59}$\lhcborcid{0000-0001-7905-0542},
C.~Abellan~Beteta$^{53}$\lhcborcid{0009-0009-0869-6798},
F.~Abudin\'en$^{61}$\lhcborcid{0000-0002-6737-3528},
T.~Ackernley$^{63}$\lhcborcid{0000-0002-5951-3498},
A.A.~Adefisoye$^{71}$\lhcborcid{0000-0003-2448-1550},
B.~Adeva$^{49}$\lhcborcid{0000-0001-9756-3712},
M.~Adinolfi$^{57}$\lhcborcid{0000-0002-1326-1264},
P.~Adlarson$^{87,44}$\lhcborcid{0000-0001-6280-3851}
% ...

{\footnotesize \it

% ...
$^{39}$Nikhef National Institute for Subatomic Physics, Amsterdam, Netherlands\\
% ...
$^{44}$National Center for Nuclear Research (NCBJ), Warsaw, Poland\\
% ...
$^{49}$Instituto Galego de F{\'\i}sica de Altas Enerx{\'\i}as (IGFAE), Universidade de Santiago de Compostela, Santiago de Compostela, Spain\\
% ...
$^{53}$Physik-Institut, Universit{\"a}t Z{\"u}rich, Z{\"u}rich, Switzerland\\
% ...
$^{57}$H.H. Wills Physics Laboratory, University of Bristol, Bristol, United Kingdom\\
% ...
$^{59}$Department of Physics, University of Warwick, Coventry, United Kingdom\\
% ...
$^{61}$School of Physics and Astronomy, University of Edinburgh, Edinburgh, United Kingdom\\
$^{62}$School of Physics and Astronomy, University of Glasgow, Glasgow, United Kingdom\\
$^{63}$Oliver Lodge Laboratory, University of Liverpool, Liverpool, United Kingdom\\
% ...
$^{71}$Syracuse University, Syracuse, NY, United States\\
% ...
$^{87}$Department of Physics and Astronomy, Uppsala University, Uppsala, Sweden, associated to $^{62}$\\
% ...
}
\end{flushleft}

\end{document}